\documentclass[twocolumn,aps]{revtex4}
\usepackage{amssymb}
\usepackage{amsmath}
\usepackage{graphicx}
\usepackage{lscape}
\usepackage{booktabs}
\usepackage{multirow}
\begin{document}
\title{Predictions for production of $\rm{^3_\Lambda H}$ and $\rm{{^3_{\overline \Lambda}\overline H}}$
in isobaric $^{96}_{44}$Ru+$^{96}_{44}$Ru and $^{96}_{40}$Zr+$^{96}_{40}$Zr
collisions at $\sqrt{s_{\rm{NN}}}$ = 200 GeV}
\author{Zhi-Lei She$^{1,2}$,  Gang Chen$^{2,} $\footnote{Corresponding Author:
chengang1@cug.edu.cn}, Dai-Mei Zhou$^3$, Liang Zheng$^2$, Yi-Long Xie$^2$, Hong-Ge Xu$^2$}
\affiliation{
  $^1$ Institute of Geophysics and Geomatics, China University of Geosciences, Wuhan,430074, China.
\\$^2$ School of Mathematics and Physics, China University of Geosciences, Wuhan,430074, China.\\
${^3}$ Institute of Particle Physics, Central China Normal University, Wuhan 430079, China.}

\begin{abstract}
 The production of $\rm{^3_\Lambda H}$ and $\rm{{^3_{\overline \Lambda}\overline H}}$, as well as $\rm{^3H}$, $\rm{{^3\overline H}}$, $\rm{^3He}$, and $\rm{{^3\overline {He}}}$ are studied in central collisions of isobars $^{96}_{44}$Ru+$^{96}_{44}$Ru and $^{96}_{40}$Zr+$^{96}_{40}$Zr at $\sqrt{s_{\rm{NN}}}=200$ GeV, using the dynamically constrained phase-space coalescence model and the {\footnotesize PACIAE} model with chiral magnetic effect. The yield, yield ratio, coalescence parameters, and strangeness population factor of (anti-)hypertriton
and (anti-)nuclei produced in isobaric $^{96}_{44}$Ru+$^{96}_{44}$Ru and $^{96}_{40}$Zr+$^{96}_{40}$Zr collisions are predicted. The (anti-)hypertriton and (anti-)nuclei production is found to be insensitive to the chiral magnetic effects. Experimental data of Cu+Cu, Au+Au and Pb+Pb collisions from RHIC, LHC, and the results of {\footnotesize PACIAE+DCPC} model are presented in the results for comparison.

\end{abstract}

\maketitle

\section{Introduction}
Hypernuclei and their antihypernuclei are copiously produced under conditions of extreme high temperatures and energy densities in high-energy heavy ion collisions. It creates a unique opportunity to study whether hypernuclei and antihypernuclei have the same behaviour and to investigate the difference between light (anti-)nuclei and (anti-)hypernuceli~\cite{pr760j,fp7y,npa987p}.
Hence it attracts a constant interest in studying antimatter and exploring fundamental problems in physics,
e.g., testing the fundamental CPT theorem by precisely measuring the difference of
the mass, lifetime and binding energy between hypertriton ($\rm{^3_\Lambda H}$) and its corresponding
anti-hypertriton ($\rm{{^3_{\overline \Lambda}\overline H}}$)~\cite{np16j,plb797s} in Au+Au and Pb+Pb collision systems.

The anti-hypertriton ($\rm{{^3_{\overline \Lambda}\overline H}}$), the lightest
bound antihypernucleus, consists of a antihyperon $\overline \Lambda$ , a
antiproton $\overline p$, and a antineutron $\overline n$, which had been discovered in Au+Au collisions at $\sqrt{s_{\rm{NN}}}=200$ GeV
by STAR Collaboration at the BNL Relativistic Heavy Ion Collider (RHIC)~\cite{sci328b} and then in Pb+Pb collisions
at $\sqrt{s_{\rm{NN}}}=2.76$ TeV by ALICE Collaboration at Large Hadron Collider (LHC) in CERN~\cite{plb754j}, respectively. The production of $\rm{^3_\Lambda H}$ ($\rm{{^3_{\overline \Lambda}\overline H}}$) has distinct features in heavy-ion collisions compared with corresponding normal three-body (anti-)nuclei $\rm^3{{He}}$ ($\rm ^3{\overline{He}}$) and $\rm^3{{H}}$ ($\rm ^3{\overline{H}}$),
due to the different interaction strength between hyperon-nucleon and nucleon-nucleon~\cite{epja48e}.
Their detailed production mechanism is, however, not fully understood.
Hence the related theoretical approaches on production of light (anti-)nuclei and $\rm{^3_\Lambda H}$ ($\rm{{^3_{\overline \Lambda}\overline H}}$) have been carried out in the frameworks of either the statistical thermal method~\cite{prc81v,plb697a,prc84j,prc87s,prc90s,natu561}
or the coalescence model~\cite{prc85l,prc92l,plb754n,prc93k,nst28p,prc99f}.

The existence of $\rm{^3_\Lambda H}$ ($\rm{{^3_{\overline \Lambda}\overline H}}$) in heavy-ion
reactions are observed, ranging from AGS~\cite{prc70t} up to RHIC~\cite{prc99f,sci328b} and
LHC~\cite{plb754j} collision energies, involving various collision systems, such as $\rm^{63}Cu$+$\rm^{63}Cu$,
$\rm ^{197}Au$+$\rm^{197}Au$, and $\rm^{208}Pb$+$\rm^{208}Pb$ collisions.
One can see that there exists a gap of the system size for nucleus-nucleus interactions between $\rm^{63}Cu$+$\rm^{63}Cu$ and $\rm ^{197}Au$+$\rm^{197}Au$ collisions.
However, the recent isobar program consisting of $^{96}_{44}$Ru+$^{96}_{44}$Ru and $^{96}_{40}$Zr+$^{96}_{40}$Zr collisions at the top RHIC energies of $\sqrt{s_{\rm{NN}}}$ = 200 GeV, is favored to search the presence of
Chiral Magnetic Effect (CME)~\cite{prc98y,prl121,prc99x,arx11j,ppnp107j}, and it can also be used to fill the gap
of collision system size discussed above.

The CME effect can reveal some topological and electromagnetic properties of
the quark gluon plasma(QGP) in high-energy heavy ion collisions. Charge separation is an important consequence of the CME. Ma {\emph{et al.}}~\cite{plb700gl,prc97wt,prc97lh,prc101lh} introduced an additional CME-induced charge separation to the initial conditions obtained from a multiphase transport model(AMPT)~\cite{prc72z}, to study the CME-related physics. Refs.~\cite{plb700gl,prc97wt} demonstrated that the final-state interactions can reduce the charge separation in each collision, while the relative difference of the CME signal between the two isobaric collisions is insensitive to the final-state interactions.

In this paper, the production of the final state hadrons, including $p$, $\overline p$, $\Lambda$, and $\overline \Lambda$, are simulated by the parton and hadron cascade model ({\footnotesize {PACIAE}})~\cite{cpc183b},
and an initial three-flavor dipole charge separation~\cite{plb700gl} is introduced to simulate the CME,
in $^{96}_{44}$Ru+$^{96}_{44}$Ru and $^{96}_{40}$Zr+$^{96}_{40}$Zr at the top RHIC energy of $\sqrt{s_{\rm{NN}}}$ = 200 GeV with midrapidity ($|\eta|<$ 0.5). Then, the dynamically constrained phase-space coalescence ({\footnotesize {DCPC}}) model~\cite{prc85y} is applied to study the production of $\rm{^3_\Lambda H}$($\rm{{^3_{\overline \Lambda}\overline H}}$) cluster in these two isobaric collision systems. In this study, we expect to compare and investigate the production and properties of $\rm{^3_\Lambda H}$ ($\rm{{^3_{\overline \Lambda}\overline H}}$) in $^{96}_{44}$Ru+$^{96}_{44}$Ru and $^{96}_{40}$Zr+$^{96}_{40}$Zr collision systems involving the chiral magnetic effect.

The paper is organized as follows: In sect. II, we provide a concise introduction to
the {\footnotesize PACIAE} with CME and {\footnotesize {DCPC}} model.
Sec. III contains our numerical calculations for production
and properties of $\rm{^3_\Lambda H}$ ($\rm{{^3_{\overline \Lambda}\overline H}}$).
In sec. IV, a short summary is given.

\section {MODELS}
The {\footnotesize {PACIAE}} model~\cite{cpc183b} is based on {\footnotesize {PYTHIA}} 6.4~\cite{jhep05t} and
is designed for various collision systems ranging from proton induced reactions (p+p and p+A), to nuclear
reactions (A+A). Generally, this entire model has four main physics stages composed of
the parton initiation, parton rescattering, hadronization, and hadron rescattering.

At the first stage, the nucleus-nucleus collision is decomposed into the nucleon-nucleon ($NN$) collisions
according to the collision geometry and $NN$ total cross section.
The strings created in the $NN$ collisions will break up into free partons
leading to the formation of the deconfined quark-gluon matter.
After that, the decomposed partons interact with each other relied on
the 2$\rightarrow$ 2 LO-pQCD parton-parton cross sections~\cite{plb70b}.
Here, a $K$ factor is added to account for non-perturbative QCD and higher-order corrections.
Then, the hadronization conducts via either the Lund string fragmentation model~\cite{jhep05t}
or the phenomenological coalescence model~\cite{cpc183b}.
The last step is the hadron rescattering process happening among the generated hadrons
until the hadronic freeze-out. (For more details see Ref.~\cite{cpc183b}).

To study the CME-related physics, an additional CME-induced charge separation mechanism~\cite{plb700gl},
which switches $p_{y}$ values for a fraction $f$ of the downward moving $u(\overline d)$ quarks with those
of the upward moving $\overline {u}(d)$ quarks, is needed to introduce
into the initial conditions in the original {\footnotesize {PACIAE}} model~\cite{cpc183b}.
The fraction $f$ can be described as
\begin{equation}
f=\frac{N^{+(-)}_{\uparrow(\downarrow)} - N^{+(-)}_{\downarrow(\uparrow)}}
{N^{+(-)}_{\uparrow(\downarrow)} + N^{+(-)}_{\downarrow(\uparrow)}},
\end{equation}
where $N$ denotes the number of a given quark, $+$ and $-$ represent positive and negative charges,
$\uparrow$ and $\downarrow$ are the moving directions of quarks along the $y$ axis, respectively.
As the Ref.~\cite{prc97wt} mentioned, the initial charge separation fractions $f$ are different between
isobars $^{96}_{44}$Ru+$^{96}_{44}$Ru and $^{96}_{40}$Zr+$^{96}_{40}$Zr collisions since they have a same
nucleon number but the 10\% difference in proton number. In this work, we introduce an initial three-flavor quarks($u,d,s$)
charge separation into original {\footnotesize {PACIAE}} model~\cite{cpc183b} to investigate the CME-related physics.

The {\footnotesize {DCPC}} model~\cite{prc85y} is developed to calculate production of light (anti-)nuclei
and (anti-)hypernuclei, after the final-state particles produced by {\footnotesize {PACIAE}} model~\cite{cpc183b}
in high energy collisions. Previous works of (anti-)nuclei and (anti-)hypernuclei production in different collision systems,
e.g., pp~\cite{prc85y,ijmpe23j,arx11n}, Cu+Cu~\cite{prc99f,epja55f},
Au+Au~\cite{prc86g,prc88g,jpg41g,epja54z} and Pb+Pb~\cite{epja52z,arx09z} interactions,
have been studied using the same framework.

According to the quantum statistical mechanics, one can estimate the yield of a single particle in the six-dimension
phase space by an integral
\begin{equation}
Y_1=\int_{H\leqslant E} \frac{d\vec qd\vec p}{h^3},
\end{equation}
where $H$ and $E$ represent the Hamiltonian and energy of the particle, respectively.
Similarly, the yield of N particle cluster can also be computed using the following integral
\begin{equation}
Y_N=\int ...\int_{H\leqslant E} \frac{d\vec q_1d\vec p_1...d\vec
q_Nd\vec p_N}{h^{3N}}. \label{funct1}
\end{equation}
In addition, equation~(\ref{funct1}) must meet the following constraint conditions
\begin{equation}
m_0\leqslant m_{inv}\leqslant m_0+\Delta m,
\end{equation}
\begin{equation}
|\vec q_{ij}|\leqslant D_0,(i\neq j;i,j=1,2,\ldots,N).
\end{equation}
where
\begin{equation}
m_{inv}=\Bigg[\bigg(\sum^{N}_{i=1} E_i \bigg)^2-\bigg(\sum^{N}_{i=1}
\vec p_i \bigg)^2 \Bigg]^{1/2},
\end{equation}
and $E_i$, $\vec p_i$($i$=1,2,\ldots,$N$) are the energies and momenta of particles, respectively.
$m_0$ and $D_0$ denote the rest mass and diameter of light (anti-)nuclei or (anti-)hypernuclei.
Here, the radius values $R$ = 1.74, 1.61, 5.0 fm are chosen for
$\rm^3{{He}}$ ($\rm ^3{\overline{He}}$), $\rm^3{{H}}$ ($\rm ^3{\overline{H}}$),
and $\rm{^3_\Lambda H}$ ($\rm{{^3_{\overline \Lambda}\overline H}}$)~\cite{nst28p,prc70t,ptp103h}
in this simulation, respectively. $\Delta m$ represents the allowed mass uncertainty,
and $|\vec q_{ij}|$ presents the distance between particles $i$-th and $j$-th.
The integration in Eq.~(\ref{funct1}) should be replaced by the summation
over discrete distributions, as a coarse graining process in the transport model.
\section {Results and Discussion}
At first, we can obtain the final state particles in central collisions
of isobaric Ru+Ru and Zr+Zr using the {\footnotesize PACIAE} model with {\footnotesize CME}.
This simulation works on the assumption that (anti-)hyperons heavier than $\Lambda$ ($\overline \Lambda$)
have already decayed, and the model parameters are fixed on the default values given in {\footnotesize {PYTHIA}} model, except the $K$ factor and string fragmentation parameters parj(1), parj(2), and parj(3).
These selected parameters are confirmed by roughly fitting production of
$p$ ($\overline p$) and $\Lambda$ ($\overline \Lambda$) in 0-15\% $^{96}_{44}$Ru+$^{96}_{44}$Ru
and $^{96}_{40}$Zr+$^{96}_{40}$Zr collisions to STAR data in 20-40\% centrality Au+Au collisions
at $\sqrt{s_{\rm{NN}}}$ = 200 GeV, since their mean number of participating nucleons
($\langle N_{\rm part}\rangle$) are quite similar $\sim$ 140.
Table~\ref{table1} shows the corresponding integrated yields ($dN/dy$) of $p$ and $\overline p$
with $|\eta|< 0.1$ and 0.35 $ <p_{T}< 1.2 $~GeV/c, as well as $\Lambda$ and $\overline \Lambda$ within
$|\eta|< 0.5$ and 0.5 $ < p_{T}< 8.0$~GeV/c, respectively. $p$ and $\overline p$ take into account of contributions
from primordial $\Lambda$ decays. For comparison, the STAR experimental
data of 20-40\% Au+Au collisions~\cite{prl108g,prc79b} are also presented.
It can be seen from Tab.~\ref{table1} that the yields of particles
($p$, $\overline p$, $\Lambda$, and $\overline \Lambda$) in $^{96}_{44}$Ru+$^{96}_{44}$Ru collision
are the same as those of $^{96}_{40}$Zr+$^{96}_{40}$Zr collisions at the specified collision centrality.
Moreover, the results of {\footnotesize PACIAE} model with {\footnotesize CME} for the two isobaric nuclear collisions are well consistent with the measured STAR data for Au+Au collisions. The same fitted parameters
of $K$ = 3.0, parj(1) = 0.13, parj(2) = 0.65, and parj(3) = 0.44 are chosen
for these two isobaric nuclear collision systems.

\begin{table}[htbp]
\caption{The integrated yield $dN/dy$ of particles ($p$, $\overline {p}$, $\Lambda$ and $\overline \Lambda$)
in 0-15\% centrality $^{96}_{44}$Ru+$^{96}_{44}$Ru and $^{96}_{40}$Zr+$^{96}_{40}$Zr collisions
of $\sqrt{s_{\rm{NN}}}=200$~GeV by {\footnotesize PACIAE} with {\footnotesize CME}, as compared to 20-40\% Au+Au collisions in STAR experimental
data~\cite{prc79b,prl108g}. Here $p$ ($\overline{p}$) are inclusive of contributions
from primordial $\rm \Lambda$ ($\rm \overline \Lambda$) decays.}
\setlength{\tabcolsep}{3.pt}
\renewcommand{\arraystretch}{0.9}
\begin{tabular}{cccc}
\hline  \hline
 {Particle} &\multicolumn{2}{c}{{\footnotesize PACIAE}}&\multicolumn{1}{c}{STAR} \\ \cline{2-4}
type & Ru+Ru(0-15\%) & Zr+Zr(0-15\%)&\multicolumn{1}{c}{Au+Au(20-40\%) }  \\\hline
$\langle N_{\rm part}\rangle$  &$139.5\pm 1.4$&  $139.5\pm 1.4$  &142.4 $\pm$ 5.3 \\
$p$                       &11.13 $\pm$ 0.03 &11.12 $\pm$ 0.03  &11.85 $\pm$ 1.15  \\
$\overline{p}$            &9.60 $\pm$ 0.02 &9.59 $\pm$ 0.01  & 9.33 $\pm$ 0.91  \\
$\rm \Lambda$             &5.57 $\pm$ 0.02 &5.55 $\pm$ 0.01  &5.70 $\pm$ 0.55  \\
$\rm \overline \Lambda$   &4.51 $\pm$ 0.01 &4.50 $\pm$ 0.01  &4.53 $\pm$ 0.34  \\
\hline \hline
\end{tabular} \label{table1}
\end{table}

\begin{figure}[tbp]
\includegraphics[width=0.45\textwidth]{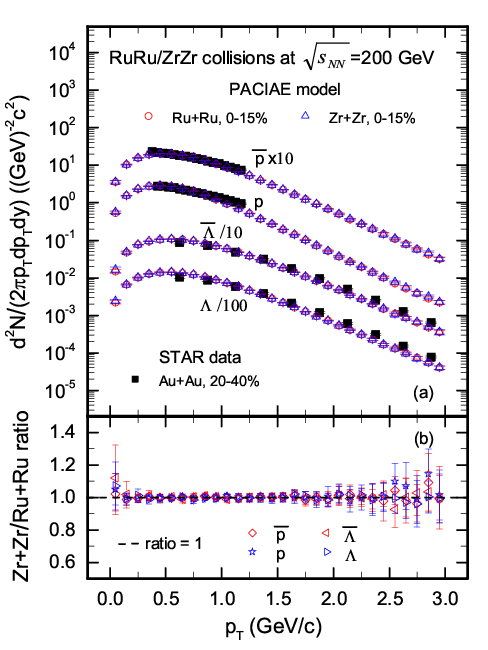}
\caption{(a) The transverse momentum spectrum of particles ( $p$, $\overline p$, $\Lambda$, $\overline \Lambda$)
in midrapidity $^{96}_{44}$Ru+$^{96}_{44}$Ru and $^{96}_{40}$Zr+$^{96}_{40}$Zr collisions at $\sqrt{s_{\rm{NN}}}=200$~GeV.
The open symbols show the results of {\footnotesize PACIAE} model with {\footnotesize CME}, and the solid symbols
show the results from STAR data~\cite{prc79b,prl108g}. For clarity the spectra data are divided by powers of 10.
(b) The yield ratios of particles ($p$, $\overline p$, $\Lambda$, $\overline \Lambda$) produced
in $^{96}_{44}$Ru+$^{96}_{44}$Ru collisions to that in $^{96}_{40}$Zr+$^{96}_{40}$Zr collisions.
} \label{tu1}
\end{figure}

Figure~\ref{tu1}(a) presents the transverse momentum distributions of $p$ ($\overline p$)
and $\Lambda$ ($\overline \Lambda$) (open symbols) in 0-15\% $^{96}_{44}$Ru+$^{96}_{44}$Ru
and $^{96}_{40}$Zr+$^{96}_{40}$Zr collisions at $\sqrt{s_{\rm{NN}}}$ = 200 GeV calculated
by the {\footnotesize PACIAE} model with {\footnotesize CME}. The STAR experimental data for 20-40\% Au+Au collisions taken
from Refs.~\cite{prc79b,prl108g} are shown by the solid symbols. It can be seen that the transverse momentum spectrum
of particles $p$ ($\overline p$) and $\Lambda$ ($\overline \Lambda$) simulated
by {\footnotesize PACIAE} model with {\footnotesize CME} are compatible with the STAR data within uncertainties.
Besides, figure~\ref{tu1}(b) shows the distribution of the invariant yield ratios of particles
for $^{96}_{44}$Ru+$^{96}_{44}$Ru and $^{96}_{40}$Zr+$^{96}_{40}$Zr collisions at $\sqrt{s_{\rm{NN}}}$ = 200 GeV,
as a function of $p_{T}$. Obviously, one can see from figure~\ref{tu1}(b) that there is no significant difference
for transverse momentum spectra of (anti-)particles between the two isobaric nuclear collisions,
except the fluctuation at higher $p_{T}$.

\begin{table}[htbp]
\caption{The integrated yields $dN/dy$ of (anti-)particles $p$ ($\overline {p}$), $\Lambda$ ($\overline \Lambda$),
and (anti-)nuclei $\rm{^3_\Lambda H}$ ($\rm{{^3_{\overline \Lambda}\overline H}}$),
$\rm^3{{He}}$ ($\rm ^3{\overline{He}}$), $\rm^3{{H}}$ ($\rm ^3{\overline{H}}$) calculated
by {\footnotesize PACIAE+DCPC} model with {\footnotesize CME} in 0-10\% $^{96}_{44}$Ru+$^{96}_{44}$Ru and $^{96}_{40}$Zr+$^{96}_{40}$Zr
collisions of $\sqrt{s_{\rm{NN}}}=200$~GeV with $|\eta|<0.5$. Here $p$ ($\overline{p}$) productions
from $\rm \Lambda$ ($\rm \overline \Lambda$) feed down contribution is excluded in the coalescence procedure.}
\setlength{\tabcolsep}{11.pt}
\renewcommand{\arraystretch}{1.1}
\begin{tabular}{ccc}
\hline  \hline
Nucleus type & Ru+Ru(0-10\%) &Zr+Zr(0-10\%)  \\ \hline
$\langle N_{\rm part}\rangle$  &$151.8\pm 1.4$ &  $151.8\pm 1.4$  \\
$p$                            &8.03 $\pm$ 0.02 &8.02 $\pm$ 0.02       \\
$\overline{p}$                 &6.51 $\pm$ 0.01 &6.50 $\pm$ 0.03       \\
$\rm \Lambda$                  &7.04 $\pm$ 0.01 &7.03 $\pm$ 0.01       \\
$\rm \overline \Lambda$        &5.64 $\pm$ 0.01 &5.65 $\pm$ 0.01       \\
$\rm{^3_\Lambda H}$ ($10^{-5}$)                         &6.50 $\pm$ 0.06  &6.45 $\pm$ 0.07 \\
$\rm{{^3_{\overline \Lambda}\overline H}}$ ($10^{-5}$)  &3.15 $\pm$ 0.03  &3.12 $\pm$ 0.03 \\
$\rm^3{{He}}$ ($10^{-5}$)                               &8.57 $\pm$ 0.09  &8.38 $\pm$ 0.08 \\
$\rm ^3{\overline{He}}$ ($10^{-5}$)                     &4.41 $\pm$ 0.04  &4.32 $\pm$ 0.04 \\
$\rm^3{{H}}$ ($10^{-5}$)                                &8.61 $\pm$ 0.08  &8.51 $\pm$ 0.09\\
$\rm ^3{\overline{H}}$ ($10^{-5}$)                      &4.79 $\pm$ 0.04  &4.73 $\pm$ 0.04 \\
\hline \hline
\end{tabular} \label{table2}
\end{table}

In the following, we generate $4.0 \times 10^8$ most central (0-10\%) events for $^{96}_{44}$Ru+$^{96}_{44}$Ru
and $^{96}_{40}$Zr+$^{96}_{40}$Zr collisions at $\sqrt{s_{\rm{NN}}}=200$~GeV using
the {\footnotesize PACIAE} model with {\footnotesize CME}, respectively. These (anti-)nucleons
and (anti-)hyperons produced within {\footnotesize PACIAE} with {\footnotesize CME} are used
as input for the {\footnotesize DCPC} model. The proton productions from $\rm \Lambda$ feed down contribution
is excluded in the coalescence procedure. Then, we obtain the integrated yields $dN/dy$ of
light (anti-)nuclei and (anti-)hypertriton nuclei with $|\eta|< 0.5$ and $p_{T}< 3.0$~GeV/c for
the most central bin of 0-10\%, respectively. Here we choose the parameter $\Delta m = 1.53$ MeV
for $\rm{^3_\Lambda H}$ ($\rm{{^3_{\overline \Lambda}\overline H}}$), and $\Delta m = 2.13$ MeV
for $\rm^3{{He}}$ ($\rm ^3{\overline{He}}$) and $\rm^3{{H}}$ ($\rm ^3{\overline{H}}$).

Table~\ref{table2} presents the integrated yields $dN/dy$ of (anti-)hyperons and
(anti-)hypertriton ($\Lambda, \overline \Lambda$, $\rm{^3_\Lambda H}$, $\rm{{^3_{\overline \Lambda}\overline H}}$),
as well as (anti-)nuclei ($p, \overline {p}$, $\rm^3{{He}}, \rm ^3{\overline{He}}$, $\rm^3{{H}}, \rm ^3{\overline{H}}$)
calculated by the {\footnotesize PACIAE}+{\footnotesize DCPC} model with {\footnotesize CME}
in most central (0-10\%) $^{96}_{44}$Ru+$^{96}_{44}$Ru and $^{96}_{40}$Zr+$^{96}_{40}$Zr collisions
at $\sqrt{s_{\rm{NN}}}=200$~GeV, respectively. It can be seen that the yields of (anti-)hypertriton,
(anti-)tritium, and (anti-)helium-3 nuclei in central $^{96}_{44}$Ru+$^{96}_{44}$Ru and $^{96}_{40}$Zr+$^{96}_{40}$Zr collisions at $\sqrt{s_{\rm{NN}}}=200$~GeV from {\footnotesize PACIAE}+{\footnotesize DCPC} simulations, are all at the order of $10^{-5}$. However, the yields of (anti-)hypernuclei are less than that of corresponding (anti-)nuclei with the equal baryon numbers. The yields of (anti-)hypernuclei and (anti-)nuclei
in isobaric $^{96}_{44}$Ru+$^{96}_{44}$Ru and $^{96}_{40}$Zr+$^{96}_{40}$Zr collisions are the same
within the range of uncertainty, indicating that the production of (anti-)hypernuclei and
light (anti-)nuclei between the two isobaric nuclei-nuclei collisions is insensitive to the difference in charge.

\begin{table*}[htbp]
\caption{The (anti-)nucleus ratios from the {\footnotesize PACIAE}+{\footnotesize DCPC} model with {\footnotesize CME} in central (0-10\%) $^{96}_{44}$Ru+$^{96}_{44}$Ru and $^{96}_{40}$Zr+$^{96}_{40}$Zr collisions
at $\sqrt{s_{\rm{NN}}}=200$~GeV. The top section of the table shows the three ratios of anti-nucleus
to nucleus, followed by the mixed ratios of (anti-)nucleus to (anti-)nucleus. The ratios between proton,
anti-proton, hyperon, and anti-hyperon are shown at the bottom. STAR data are taken from Cu+Cu and Au+Au collisions
at $\sqrt{s_{\rm{NN}}}=200$~GeV~\cite{sci328b,prc79b,prl108g,prc83m,phd09j}, respectively.}
\setlength{\tabcolsep}{25.pt}
\renewcommand{\arraystretch}{0.9}
\begin{tabular}{cccccc} \hline  \hline
Nucleus &\multicolumn{2}{c}{\footnotesize STAR} &\multicolumn{2}{c}{\footnotesize PACIAE+DCPC}\\

\cmidrule[0.3pt](l{0.01cm}r{0.15cm}){2-3}
\cmidrule[0.3pt](l{0.15cm}r{0.01cm}){4-5}
ratio &Cu+Cu &Au+Au &Ru+Ru &Zr+Zr \\\hline
$\rm{{^3_{\overline \Lambda}\overline H}/{^3_{\Lambda}H}}$ &  $-$  & ${0.49\pm0.18\pm0.07}$    &${0.48\pm0.02}$       &${0.48\pm0.01}$ \\
$\rm{{^3{\overline{He}}}/{^3{He}}}$                        &  $0.46\pm 0.17$  & ${0.45\pm0.02\pm0.04}$    &${0.51\pm0.01}$       &${0.51\pm0.01}$ \\
$\rm{{^3{\overline H}}/{^3H}}$                             &  $-$  & $-$                       &${0.56\pm0.01}$       &${0.56\pm0.01}$ \\
$\rm{{^3_{\overline \Lambda}\overline H}/{^3{\overline{He}}}}$ & $-$  & ${0.89\pm0.28\pm0.13}$ &${0.71\pm0.02}$   &${0.72\pm0.01}$ \\
$\rm{{^3_{\Lambda}H}/{^3{He}}}$                                & $-$  & ${0.82\pm0.16\pm0.12}$ &${0.76\pm0.02}$   &${0.77\pm0.01}$ \\
$\rm{{^3_{\overline \Lambda}\overline H}/{^3{\overline H}}}$   & $-$  & $-$                    &${0.66\pm0.02}$   &${0.66\pm0.01}$ \\
$\rm{{^3_{\Lambda}H}/{^3{H}}}$                                 & $-$  & $-$                    &${0.75\pm0.02}$   &${0.76\pm0.02}$ \\
$\rm{\overline p /p}$                  & 0.80$\pm$0.04   & 0.79$\pm$0.11   & $0.81\pm0.01$  &  $0.81\pm0.01$ \\
$\overline \Lambda /\Lambda$           & 0.82$\pm$0.12   & 0.80$\pm$0.10   & $0.80\pm0.01$  &  $0.80\pm0.01$   \\
$\Lambda /\rm p$                       & 0.84$\pm$0.09   &  $-$            & $0.88\pm0.01$  &  $0.88\pm0.01$ \\
$\overline \Lambda /\rm\overline p$    & 0.83$\pm$0.08   &  $-$            & $0.87\pm0.01$  &  $0.87\pm0.01$   \\
\hline \hline
\\\end{tabular} \label{table3}
\end{table*}

In order to understand the fundamental properties of antimatter in nuclear collisions, we provide a systematic
investigation to the yield ratios of different (anti-)nuclei and (anti-)hypernuclei, which are deeply related to
the fractions of constituent nucleons in the naive coalescence framework~\cite{prc84j,prc85l}.
For instance, the yield ratio of $\rm{{^3_{\overline \Lambda}\overline H}}/\rm{^3_\Lambda H}$  should be proportional to $(\overline p/p)(\overline n/n)(\overline \Lambda/\Lambda)$, which is approximate to $(\overline p/p)^2(\overline \Lambda/\Lambda)$, i.e,
\begin{equation}
\frac{\rm {^3_{\overline \Lambda}\overline H}}{\rm{^3_{\Lambda}H}}=\\
 \frac{\rm {\overline p \overline n \overline\Lambda}}{\rm{pn\Lambda}}\simeq\\
 (\frac{\rm \overline p}{\rm p})^2\frac{\rm \overline\Lambda}{\Lambda}. \label{funct6}
\end{equation}

\begin{figure}[htbp]
\includegraphics[width=0.45\textwidth]{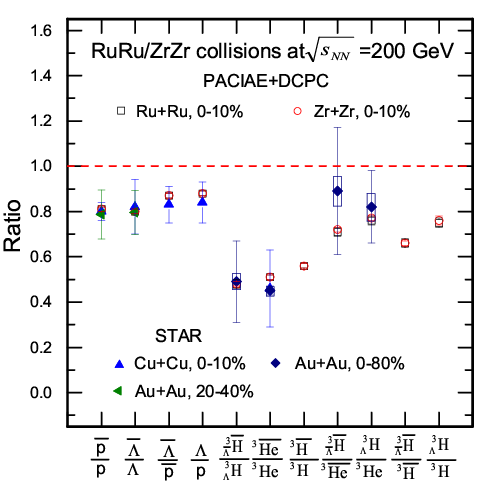}
\caption{The ratios and mixed ratios of (anti-)matter form {\footnotesize PACIAE}+{\footnotesize DCPC} model with {\footnotesize CME} (open symbols) in 0-10\% Ru+Ru and Zr+Zr collisions, compared with Cu+Cu and Au+Au collisions.
The data are taken from STAR~\cite{sci328b,prc79b,prl108g,prc83m,phd09j}. The vertical lines and error boxes
show statistical and systematic errors, respectively.}\label{tu2}
\end{figure}

Table~\ref{table3} represents the yield ratios of antiparticles to particles ($\overline p/p$, $\overline \Lambda/\Lambda$, $\rm{{^3_{\overline \Lambda}\overline H}}/\rm{^3_\Lambda H}$, $\rm ^3{\overline{He}}/\rm^3{{He}}$, $\rm ^3{\overline{H}}/\rm^3{{H}}$), and the mixed ratios ($\overline \Lambda/\overline p$, $\Lambda/p$, $\rm{{^3_{\overline \Lambda}\overline H}}/\rm ^3{\overline{He}}$,$\rm{^3_\Lambda H}/\rm^3{{He}}$, $\rm{{^3_{\overline \Lambda}\overline H}}/\rm ^3{\overline{H}}$, $\rm{^3_\Lambda H}/\rm^3{{H}}$)
calculated by {\footnotesize PACIAE}+{\footnotesize DCPC} model with {\footnotesize CME} in $^{96}_{44}$Ru+$^{96}_{44}$Ru and $^{96}_{40}$Zr+$^{96}_{40}$Zr collisions at $\sqrt{s_{\rm{NN}}}=200$~GeV. One can see from Table~\ref{table3} that the yield ratios of $\rm{{^3_{\overline \Lambda}\overline H}/{^3_{\Lambda}H}}$,
$\rm{{^3{\overline{He}}}/{^3{He}}}$, and $\rm{{^3{\overline H}}/{^3H}}$ are the same within the error range,
although their yields are not the same as shown in Table~\ref{table2}.
And the mixed ratio values of (anti-)hypernuclei to (anti-)nuclei($\rm{{^3_{\overline \Lambda}\overline H}}/\rm ^3{\overline{He}}$, $\rm{^3_\Lambda H}/\rm^3{{He}}$, $\rm{{^3_{\overline \Lambda}\overline H}}/\rm ^3{\overline{H}}$, $\rm{^3_\Lambda H}/\rm^3{{H}}$) are also the same in the range of uncertainty in central isobaric Ru+Ru and Zr+Zr at $\sqrt{s_{\rm{NN}}}=200$~GeV.
Besides, the yield ratio results of antiparticles to particles and their mixed ratios simulated
by {\footnotesize PACIAE}+{\footnotesize DCPC} model with {\footnotesize CME} are found to be
in agreement with the above theoretical interpretation within uncertainties.

Fig.~\ref{tu2} and Table~\ref{table3} show that the ratios of anti-nuclei to nuclei ($\rm{{^3_{\overline \Lambda}\overline H}}/\rm{^3_\Lambda H}$, $\rm^3{\overline{He}}/\rm^3{{He}}$, $\rm ^3{\overline{H}}/\rm^3{{H}}$) are less than 1, meaning that the yields of antiparticles is less than that of corresponding particles; similarly, the mixed ratio values indicate that the yields of the (anti-)hypertriton are less than that of (anti-)nuclei. Our simulation results are consistent with the STAR data of Cu+Cu~\cite{prl108g,prc83m,phd09j}
and Au+Au~\cite{sci328b,prc79b,prl108g} collisions at $\sqrt{s_{\rm{NN}}}=200$~GeV.

\begin{figure}[htbp]
\includegraphics[width=0.45\textwidth]{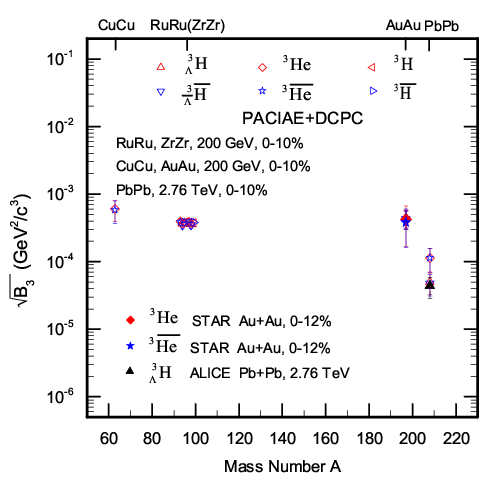}
\caption{The coalescence parameters $\sqrt{B_3}$ of $\rm{^3_\Lambda H}(\rm{{^3_{\overline \Lambda}\overline H}})$,
as well as $\rm{^3H}(\rm{{^3\overline H}})$ and $\rm{^3He}(\rm{{^3\overline {He}}})$ are compared in
isobars $^{96}_{44}$Ru+$^{96}_{44}$Ru and $^{96}_{40}$Zr+$^{96}_{40}$Zr collisions at $\sqrt{s_{\rm{NN}}}=200$ GeV
with {\footnotesize PACIAE}+{\footnotesize DCPC} model with {\footnotesize CME}. It is also compared with
the results from different collisions of Cu+Cu~\cite{epja55f}, Au+Au~\cite{epja54z}, and Pb+Pb~\cite{epja52z} collisions. The open symbols denote the results computed by {\footnotesize PACIAE}+{\footnotesize DCPC}. The solid points take from STAR~\cite{arx09b} and ALICE~\cite{plb754j}. The error bars show statistical uncertainties.}\label{tu3}
\end{figure}

In nuclear collisions, the invariant yields for production of (anti-)hypernuclei and (anti-)nuclei can be related to
the primordial yields of (anti-)nucleons in the coalescence framework~\cite{prc59r,prl37h} by Eq.~(\ref{funct7})
\begin{equation}
E_A\frac{d^3N_A}{d^3P_A} \thickapprox B_A(E_P\frac{d^3N_P}{d^3P_P})^A, \label{funct7}
\end{equation}
where $Ed^3N/d^3p$ stands for the invariant yields of nucleons or light (anti-)nuclei and (anti-)hypernuclei,
and $A$ is the atomic mass number, respectively. $B_A$ represents the coalescence parameters, which relates to
the freeze-out correlation volume, i.e., $B_A\propto V^{1-A}_{f}$. $p_A, p_p$ denote their momentum, with $p_A =A p_p$ assumed.

Fig.~\ref{tu3} presents coalescence parameters $\sqrt{B_3}$ of $\rm{^3_\Lambda H}$ ($\rm{{^3_{\overline \Lambda}\overline H}}$), as well as $\rm^3{{He}}$ ($\rm ^3{\overline{He}}$) and $\rm^3{{H}}$ ($\rm ^3{\overline{H}}$) in Ru+Ru and Zr+Zr collisions.
Meanwhile, the results from different collision systems of Cu+Cu~\cite{prc99f}, Au+Au~\cite{epja54z}, and Pb+Pb~\cite{epja52z} collisions are compared as a function of atomic mass number $A$.
One can see that $\sqrt{B_3}$ calculated by {\footnotesize PACIAE}+{\footnotesize DCPC} model with {\footnotesize CME} is constant within the error range from Cu+Cu ($A=63$) to Au+Au ($A=197$) collisions at top RHIC energy. The coalescence parameter is found to drop from RHIC energy to PbPb collisions at the LHC energy. This decreasing can be understood as the correlation volume of the QGP fireball at LHC becomes larger, as indicated by the pion HBT measurement~\cite{epja54z,arx09b}.

Specifically, with respect to 0-10\% Ru+Ru or Zr+Zr collisions, the values of $\sqrt{B_3}$ for
nuclei $\rm{^3_\Lambda H}$, $\rm^3{{He}}$, $\rm^3{{H}}$ and their corresponding anti-nuclei $\rm{{^3_{\overline \Lambda}\overline H}}$, $\rm ^3{\overline{He}}$, $\rm ^3{\overline{H}}$ are about $(3.63\pm0.46) \times10^{-4}$, $(3.90\pm0.42) \times10^{-4}$, $(3.79\pm0.39) \times10^{-4}$
and $(3.46\pm0.44) \times10^{-4}$, $(3.81\pm0.40) \times10^{-4}$, $(3.75\pm0.41) \times10^{-4}$, respectively.
It is clear that the value of $\sqrt {B_3}$ of (anti-)hypertriton is the same as that of (anti-)nuclei within the uncertainties. One can also find that the negative (hyper-)nuclei are the same as
that of positive (hyper-)nuclei. Meanwhile, the experiment data of 0-12\% Au+Au in STAR~\cite{arx09b}
and 0-10\% Pb+Pb from ALICE~\cite{plb754j} are also presented in Fig.~\ref{tu3}.

\begin{figure}[htbp]
\includegraphics[width=0.45\textwidth]{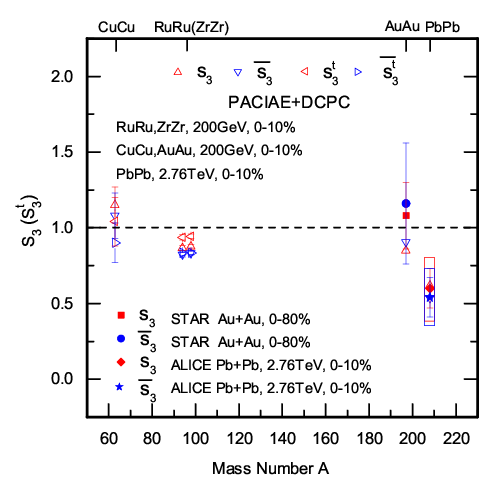}
\caption{Comparison of strangeness population factor $s_{3}(s_{3}^{t})$ in isobars $^{96}_{44}$Ru+$^{96}_{44}$Ru
and $^{96}_{40}$Zr+$^{96}_{40}$Zr collisions at $\sqrt{s_{\rm{NN}}}=200$ GeV
by {\footnotesize PACIAE}+{\footnotesize DCPC} model with {\footnotesize CME}. It is also compared with the results of Cu+Cu~\cite{prc99f}, Au+Au~\cite{epja54z}, and Pb+Pb~\cite{epja52z} collisions.
The open symbols denote the results computed by {\footnotesize PACIAE}+{\footnotesize DCPC}, and the solid points denote data from STAR~\cite{sci328b} and ALICE~\cite{plb754j}. Error bars and error boxes denote statistical and systematic errors, respectively.} \label{tu4}
\end{figure}

The strangeness population factor $s_3$, should be about one in
the coalescence model for particle production, as first suggested in Ref.~\cite{prc70t}.
It is a possible tool to study the nature of a quark-gluon plasma
created in high energy nuclear collisions~\cite{plb684s},
due to its sensitivity to the local baryon-strangeness correlation~\cite{prl95v,plb714j}.
This factor typically is written as
\begin{equation}
s_3 =(\rm{^3_\Lambda H} \times p)/(\rm^3{{He}} \times \Lambda),
\end{equation}
which can be straightforwardly extended to $\rm ^3{H}$ expressed as
\begin{equation}
s_{3}^{t}=(\rm{^3_\Lambda H} \times p)/(\rm^3{{H}} \times \Lambda).
\end{equation}

In Fig.~\ref{tu4}, we compare the values of strangeness population factor $s_3 (\overline {s_3})$
and $s_{3}^{t} (\overline {s_{3}^{t}})$ calculated by {\footnotesize PACIAE}+{\footnotesize DCPC} model with {\footnotesize CME} in central Ru+Ru and Zr+Zr collisions. Besides, the values of Cu+Cu~\cite{prc99f}, Au+Au~\cite{epja54z}, and Pb+Pb~\cite{epja52z} collisions varying with mass number A are also presented in this figure. It is shown that the values of $s_3 (\overline {s_3})$ and $s_{3}^{t} (\overline {s_{3}^{t}})$ for three-body coalescence are all the same within the error range as $A$ increases from 63 to 197 in central (0-10\%) Cu+Cu, Ru+Ru(Zr+Zr) to Au+Au collisions at RHIC energy. But the values of $s_3 (\overline {s_3})$ and $s_{3}^{t} (\overline {s_{3}^{t}})$ decrease in Pb+Pb collisions at $\sqrt{s_{\rm{NN}}}=2.76$ TeV. This may be interpreted as the (anti-)$\Lambda$ particles freezeout earlier than (anti-)nucleons but their relative freeze-out time is closer at LHC than at RHIC~\cite{prc93k}.

Numerically, the present values of $s_3, \overline s_{3}$ and ${s_3^t}, \overline {s_{3}^{t}}$ are
$0.87\pm 0.02, 0.82\pm 0.02$ and $0.94\pm 0.02, 0.83\pm 0.02$ in 0-10\% Ru+Ru collisions,
and $0.88\pm 0.01, 0.83\pm 0.01$ and $0.94\pm 0.02, 0.83\pm 0.02$ in 0-10\% Zr+Zr collisions, respectively.
Meanwhile, the values of $s_3(\overline {s_3})$ and $s_{3}^{t}(\overline {s_{3}^{t}})$ shown in the Fig.~\ref{tu4}
for Au+Au and Pb+Pb collisions by {\footnotesize PACIAE+DCPC} model are in agreement with
the corresponding available data from STAR~\cite{sci328b} and ALICE~\cite{plb754j} within uncertainties.

\section {Conclusion}
In this paper, we use the {\footnotesize PACIAE} model with CME
and {\footnotesize DCPC} model to simulate production of $\rm{^3_\Lambda H}$ and
$\rm{{^3_{\overline \Lambda}\overline H}}$, as well as $\rm{^3H}$, $\rm{{^3\overline H}}$, $\rm{^3He}$,
and $\rm{{^3\overline {He}}}$ in isobaric $^{96}_{44}$Ru+$^{96}_{44}$Ru and $^{96}_{40}$Zr+$^{96}_{40}$Zr
central collisions at $\sqrt{s_{\rm{NN}}} = 200$~GeV with $|\eta|<0.5$ and $p_{T}<$ 3.0 GeV/c, respectively.
We predict the yield, yield ratio, coalescence parameters, and strangeness population factor
of (anti-)hypertriton ($\rm{^3_\Lambda H}, \rm{{^3_{\overline \Lambda}\overline H}}$) and
(anti-)nuclei ($\rm{^3H}$, $\rm{{^3\overline H}}$, $\rm{^3He}, \rm{{^3\overline {He}}}$)
in isobaric Ru+Ru and Zr+Zr collisions. Then the chiral magnetic effect on the generation of (anti-)hypertriton ($\rm{^3_\Lambda H}, \rm{{^3_{\overline \Lambda}\overline H}}$) and (anti-)nuclei ($\rm{^3H}$, $\rm{{^3\overline H}}$, $\rm{^3He}, \rm{{^3\overline {He}}}$) in high-energy collisions are studied.
It is found that there is no clear difference for the generation and properties of
(anti-)hypertriton ($\rm{^3_\Lambda H}, \rm{{^3_{\overline \Lambda}\overline H}}$) and light (anti-)nuclei ($\rm{^3H}$, $\rm{{^3\overline H}}$, $\rm{^3He}, \rm{{^3\overline {He}}}$) in isobaric Ru+Ru and Zr+Zr collision systems, although these two collision systems have different CME.

In addition, the coalescence parameters $\sqrt{B_3}$ and the strangeness population factor $s_{3}(s_{3}^{t})$ that produce $\rm{^3_\Lambda H}(\rm{{^3_{\overline \Lambda}\overline H}})$, as well as $\rm{^3H}(\rm{{^3\overline H}})$ and $\rm{^3He}(\rm{{^3\overline {He}}})$ in $^{96}_{44}$Ru+$^{96}_{44}$Ru and $^{96}_{40}$Zr+$^{96}_{40}$Zr collisions are compared with those of Cu+Cu, Au+Au, and Pb+Pb collisions at $\sqrt{s_{\rm{NN}}}=200$ GeV by the {\footnotesize PACIAE} + {\footnotesize DCPC} model. One can see that the $\sqrt{B_3}$ and $s_{3}(s_{3}^{t})$ are all the same within the error range as atomic mass number $A$ increases from 63 to 197 in Cu+Cu, Ru+Ru(Zr+Zr) to Au+Au collisions at RHIC energy. But the values of $s_3 (\overline {s_3})$ and $s_{3}^{t} (\overline {s_{3}^{t}})$ decrease in Pb+Pb collisions at $\sqrt{s_{\rm{NN}}}=2.76$ TeV. This may be interpreted as the $\Lambda (\overline \Lambda)$ particles freezeout earlier than (anti-)nucleons but their relative freeze-out time is closer at LHC than at RHIC. The experimental data of Cu+Cu, Au+Au and Pb+Pb collisions from RHIC,
LHC are included in the comparison, which shows that our simulation results for the $\sqrt{B_3}$ and $s_3 (s_{3}^{t})$ of $\rm{^3_\Lambda H}$ ($\rm{{^3_{\overline \Lambda}\overline H}}$), $\rm^3{{He}}$ ($\rm ^3{\overline{He}}$), $\rm^3{{H}}$ ($\rm ^3{\overline{H}}$) are consistent with the experimental results within the error range.

\begin{center} {ACKNOWLEDGMENT} \end{center}
This work was supported by NSFC(11475149, 11775094, 11905188), as well as supported by the high-performance computing platform
of China University of Geosciences. The authors thank Dr. Feng-Xian Liu for fruitful discussions.

\end{document}